\documentclass{article}

\usepackage{amsmath,amssymb,amsthm}
\usepackage{graphicx}
\usepackage{caption}
\usepackage{subcaption}
\usepackage{url}
\usepackage{a4wide}

\theoremstyle{definition}

\begin{document}

\title{Viral marketing as epidemiological model\thanks{This is a preprint of a paper whose
final and definite form is in Proceedings of the 15th International Conference
on Computational and Mathematical Methods
in Science and Engineering.
Please cite this paper as: \emph{Helena Sofia Rodrigues and Manuel José Fonseca. Viral marketing as epidemiological model, Proceedings of the 15th International Conference
on Computational and Mathematical Methods
in Science and Engineering,
2015, 946--955.}} }

\author{Helena Sofia Rodrigues$^{1,2}$\\
{\tt \small sofiarodrigues@esce.ipvc.pt}
 \and Manuel Fonseca$^{1,3}$\\
{\tt \small manuelfonseca@esce.ipvc.pt}
}

% ----------

\date{$^1$ \mbox Business School, Viana do Castelo Polytechnic Institute,\\
Portugal\\[0.3cm]
$^2${\text{Center for Research and Development in Mathematics and Applications (CIDMA)}},
University of Aveiro, Portugal\\[0.3cm]
$^3$ Applied Management Research Unit (UNIAG)\\
APNOR, Portugal\\[0.3cm]
}

\maketitle

% ----------

%-----------------------------------------------------------------------------------------

\begin{abstract}
In epidemiology, an epidemic is defined as the spread of an infectious
disease to a large number of people in a given population within a short
period of time. In the marketing context, a message
is viral when it is broadly sent and received by the target market through
person-to-person transmission. This specific marketing
communication strategy is commonly referred as viral marketing.
Due to this similarity between an epidemic and the viral marketing process and
because the understanding of the critical factors to this communications strategy
effectiveness remain largely unknown, the mathematical models in epidemiology are
presented in this marketing specific field.
In this paper, an epidemiological model SIR (Susceptible-Infected-Recovered) to
study the effects of a viral marketing strategy is presented. It is made a comparison
between the disease parameters and the marketing application, and simulations using
the Matlab software are performed.
Finally, some conclusions are given and
their marketing implications are exposed: interactions across the parameters are found
that appear to suggest some recommendations to marketers, as the profitability
of the investment or the need to improve the targeting criteria of the
communications campaigns.

\noindent \textbf{Keywords:} viral marketing, word-of-mouth, epidemiological model,
numerical si\-mulations, infectivity, recovery rate, seed population
\smallskip

\noindent \textbf{2010 Mathematics Subject Classification:}34A34;  92D30; 91F99
\end{abstract}

%
%  Main text of the article.
%

\section{Introduction}

Viral marketing (VM) is a recent approaching markets and communication
with customers and can potentially reach a large and fast audience
\cite{Lans2010}. VM exploits existing social networks by encouraging
people to share product information
and campaigns with their friends, through email or social networks.
This type of communication has more impact in the customer, because
the information was recommended by friends and peer networks,
instead of a standard companies. Marketing campaigns that leverage
viral processes of dissemination are considered widely. They have
certain advantages over traditional mass media campaigns,
especially with the cost effectiveness issues and the ability to
reach specific consumer groups \cite{Bruyn2008}.

When a marketing message goes viral, it is analogous to an epidemic,
since involves a person-to-person transmission, spreading within a population.
Using insights from epidemiology to describe the spread of viruses, a mathematical
model of the viral marketing process is proposed. The main aims of this study is to
develop a set of simulation experiments to explore the influence of various
controlled and external factors that could influence viral campaigns.

The structure of this paper is as follows. In Section 2 a theoretical framework related
to viral marketing is exposed, and it is explained the transmission process of a marketing
campaign that turns viral. Then, an epidemiological model is proposed in Section 3,
that reflects the previous theoretical concepts. An interpretation of the mathematical model
in the marketing context is presented.
The relationship between epidemiology, mathematical modeling and
computational tools allows to build and test theories on the development
of a viral message; thus, the numerical simulations using the model
are done in Section 4, as well as the analysis of the graphics obtained.
Finally, conclusions are presented in Section 5 and some future work is proposed.

\section{Viral Marketing - Theoretical Framework}

It is well known that in recent decades, societies have undergone numerous
 formal and conceptual changes in their economic, social, political,
 communication, among others, standards as a result of continuous
 technological innovation that occurs in the most diverse fields.
 It is in this context that the process of digitization of the
 communications media assume itself as a significant change engine,
 both in terms of interpersonal relations and marketing communications.

Currently, in this context, a co-existence
between the media classified as online and offline has been witnessing, which results in
a hybridization of languages, strategies and techniques of communication,
translated into the key concepts of interaction and bidirectionality
\cite{Lavigne2002}. Under this relation the production and dissemination
of information ceased to be the monopoly of a specific entity to take
a collaborative dimension, where all individuals are connected to virtual
networks, being able to produce and share content, having the
possibility of its direct manipulation \cite{Castells2004}.

Given the ineffectiveness of traditional promotional communication
strategies - from massified and essentially unidirectional nature -
communicators entities have responded with new approaches, such as
viral marketing, which aims to minimize the resistance of consumers
to the promotional messages: establishing a parallelism between the
biological process transmission of a virus between individuals in a
real context, and the process of transmitting a message from an
internet user to another in digital context \cite{Nail2005}.

Viral marketing, in its strategic line, is directly related to the
variable promotion of the marketing mix, assuming as an alternative
communication technique, embodied in a dissemination process of
promotional content similar to the logic of a viral epidemic.
The spread of the message on a large scale is made possible by
the collaborative action of individuals in virtual networks
\cite{Barichell2010}.

Also known as Internet Word-of-mouth (WOM) marketing \cite{Woerdl2008}
viral marketing has been gaining more and more fans, from
professionals to researchers, as an alternative strategy to
traditional communication. In this context a divestment
in the traditional media by major advertisers is been observed, and the transfer of
funds to the online marketing actions is increasing \cite{Hinz2011}.
Among companies that use this communication strategy, stand out examples
such as Procter \& Gamble, Microsoft, BMW and Samsung, who take on the
viral marketing as a consumer-initiated activity that spreads the
marketing message unaltered across the market or segment in a limited
period team, mimicking an epidemic \cite{Gardner2013}.

In this context, the brands are transferring to consumers the power to
communicate and disseminate their promotional messages along its network
of contacts, with a smaller investment and a much higher velocity compared
to the traditional media, either the Internet or on mobile context.
Woerdl \cite{Woerdl2008} has rank as potential benefits of viral marketing:
inexpensive; reaches audiences within a short period of time; fast and
exponential diffusion; voluntary transmission by sender; more effective
targeting; access to diverse audience through social contacts. Concerning
to the risks associated with viral marketing, the authors stress:
uncontrollable nature, in particular loss over content and audience
reach, few possibilities to measure success and timing; negative WOM
leading to boycott, ruin, unfavorable attitudes; consumers unwilling
to provide referrals unless there is some return; legal emerging issues
have to be considered; consumers may feel exploited, cheated, used.

As key attributes of a successful viral message, we can highlight
features like entertainment, humor, curiosity, surprise, useful
information and relevant content. The more engaging and compelling
is the message, the more likely to spread successfully through a
network of contacts in continuous growth \cite{Dobele2007}.

There are countless possibilities of spread of a viral content.
This can be shared through video sharing sites, personal blogs,
groups or discussion forums, or simply being sent by email, as
web page hyperlink , or as attached content. Regarding the profile
of users that can enhance the spread, there is evidence that internet
users, who are more individualistic and/or more altruistic, trend
forward to more online content than others \cite{Ho2010}.

\section{Epidemiological Model}

The analysis of a VM campaign can be explained by a standard epidemic model
\cite{Leskovec2007, Sohn2013}.

In this paper, it is presented a \emph{SIR} model:
\begin{quote}
\begin{tabular}{ll}
$S(t)$ & --- susceptible (individuals who can contract the disease);\\
$I(t)$ & --- infected (individuals who can transmit the disease);\\
$R(t)$ & --- recovered (individuals who have been infected and have recovered).
\end{tabular}
\end{quote}

These compartments are mutually-exclusive. In this paper, it is assumed that the total population is constant,
which means that $N=S(t)+I(t)+R(t)$. To describe the model is also necessary to present a set of parameters:
\begin{quote}
\begin{tabular}{ll}
$\delta$ & --- contact rate among individuals per period of time;\\
$\tau $& --- probability of a contact between a susceptible and an infected that results \\
&  in disease transmission;\\
$\beta$ & --- infectivity;\\
$\gamma$ & --- recovery rate.
\end{tabular}
\end{quote}

The number of infections among the susceptible population in a period of time
depends on the constant rate ($\delta$) at a given period and the transmissibility ($\tau$)
of the disease given contact. So, the constant $\beta=\delta \tau$ is called the infectivitiy parameter.

The system of differential equations that translates the dynamics of VM is composed by:
\begin{equation}
\label{ode}
\begin{tabular}{l}
$
\left\{
\begin{array}{l}
\displaystyle\frac{dS(t)}{dt}
= -\beta \frac{S(t) I(t)}{N}\\
\displaystyle\frac{dI(t)}{dt} = \beta \frac{S(t) I(t)}{N}-\gamma I(t)\\
\displaystyle\frac{dR_h(t)}{dt} = \gamma I(t)
\end{array}
\right. $\\
\end{tabular}
\end{equation}
and subject to initial conditions
\begin{center}
\begin{tabular}{l}
\label{initial_conditions}
$S(0)=S_{0}, \quad  I(0)=I_{0}, \quad R(0)=R_{0}.$
\end{tabular}
\end{center}

\begin{figure}
\centering
\includegraphics[scale=0.5]{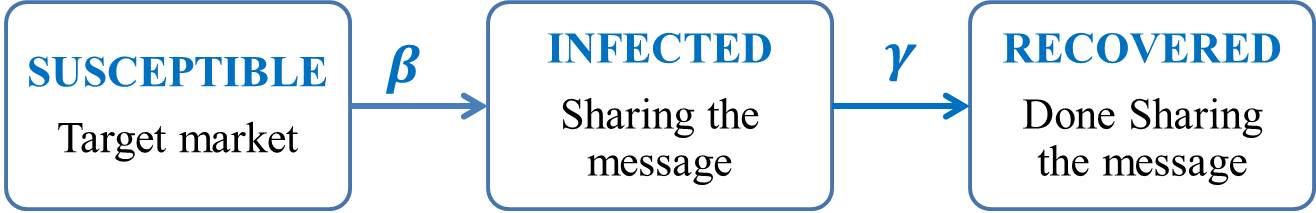}
{\caption{\label{epidemiological_model}  SIR epidemiological model for viral marketing}}
\end{figure}

The scheme of the model is shown in Figure \ref{epidemiological_model}. An arrow pointing into a compartment
is associated with a positive member of the corresponding differential equation while an arrow
pointing out of the compartment represents a negative member of the equation.

The model provides a simple and intuitive approach, modeling the viral
marketing process through epidemiological point of view, where a
disease is spread by person to person contact.
In the marketing context, a susceptible individual is a potential consumer
who may accept the message or
use an offering from a company; this compartment is called the target audience.
An infected individual passes the message or uses a unique product from
the company and/or recommend it. $\beta$ is the probability of moving
from the target audience to infective. The infected individuals start to spread
the message through social contacts. When an individual stops to share a message,
passes to the recovered compartment \cite{Gardner2013, Sohn2013}.

These parameters can also be explained in the marketing context. Infectivity ($\beta$)
is influenced by transmissibility ($\tau$) - through the marketer's efforts to seed the market
with the message and the
acceptability of the cost in time and effort to pass along the message - and the contact rate ($\delta$),
where the extent and the suitability of the social network are key aspects \cite{Gardner2013, Marutschke2014}.
Here, an target member, must perceived the message, have predisposition to deal with the message and finally,
must be motivated to share it. Factors such as exhausting mailing lists, forgetting,
no more interested by this message or
divergent attentions can leave to an individual stop to sharing the message at a rate of $\gamma$,
becoming recovered from the VM campaign. If the exit (recovery) rate is high,
the infective recovers rather quickly.

Thus, it is important to analyze and to assess the impact of a message to the target audience
and the number of members of the target market who have actively shared the message.

\section{Simulations}

The software used in the simulations was \texttt{Matlab}, with the routine \texttt{ode45}.
This solver is based on an explicit Runge-Kutta (4,5) formula, the Dormand-Prince pair.
That means the numerical solver ode45 combines fourth and fifth order methods, both of
which are similar to the classical fourth order Runge-Kutta method. These vary the step
size, choosing it at each step an attempt to achieve the desired accuracy.

\subsection{Parameter Effects}

In this section we made simulations with system (\ref{ode}),
using the following initial values for differential equations:

\begin{center}
\begin{tabular}{l}
$S(0)=900, \quad  I(0)=100, \quad R(0)=0$.\\
\end{tabular}
\end{center}

The analysis of the parameters variation will be focus on the infect compartment -
to visualize the transmission shape - and also on the number of recovered individuals
 at final time, since this population
captures the cumulative number of sharing individuals. The analysis will be done in
dimensionless time, because some campaigns could came viral in days and other in hours or minutes.

\begin{figure}
\centering
\begin{subfigure}[b]{0.45\textwidth}
\centering
\includegraphics[scale=0.45]{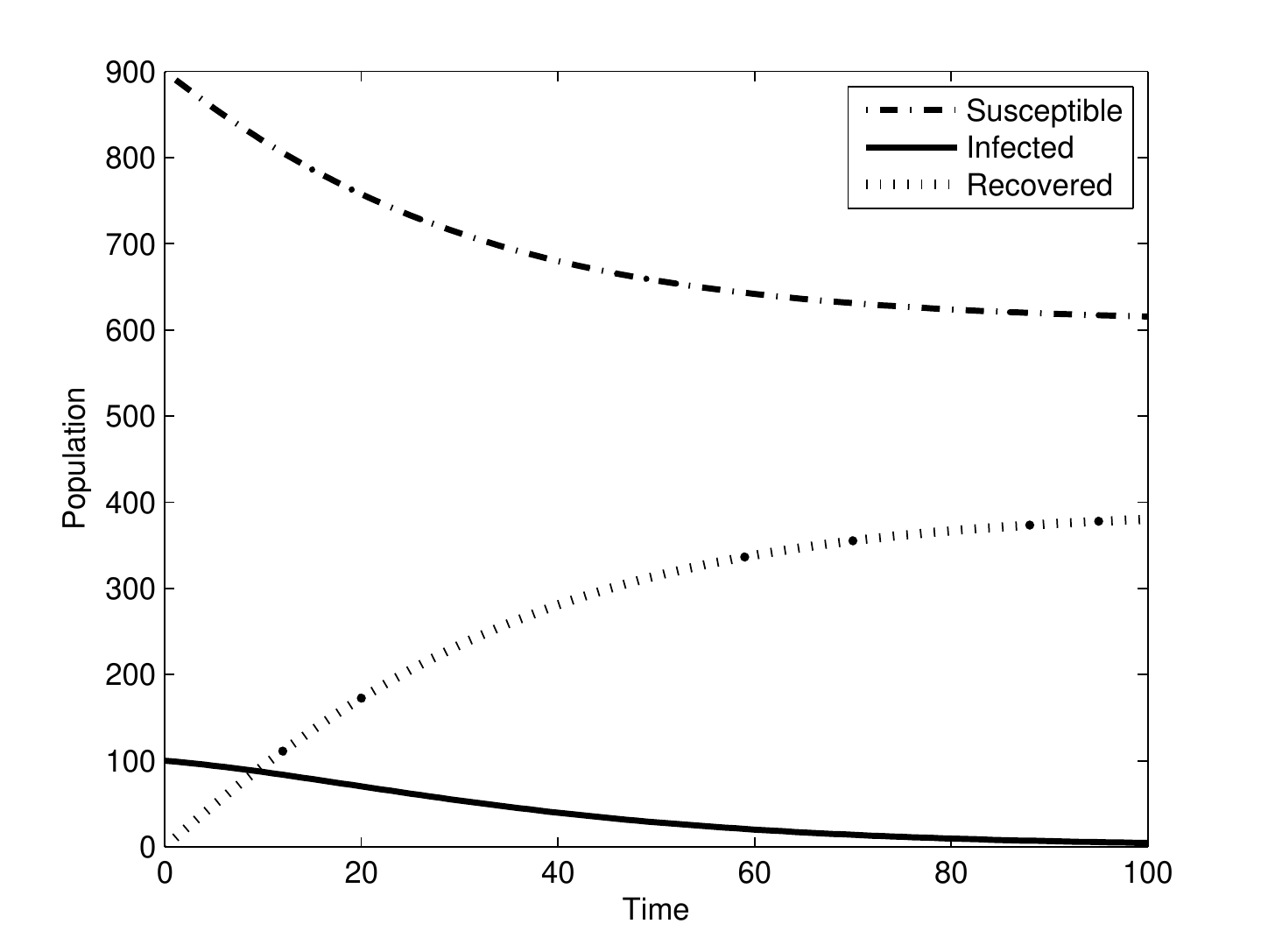}
\caption{$\beta=0.1$}
\end{subfigure}%
\begin{subfigure}[b]{0.45\textwidth}
\centering
\includegraphics[scale=0.45]{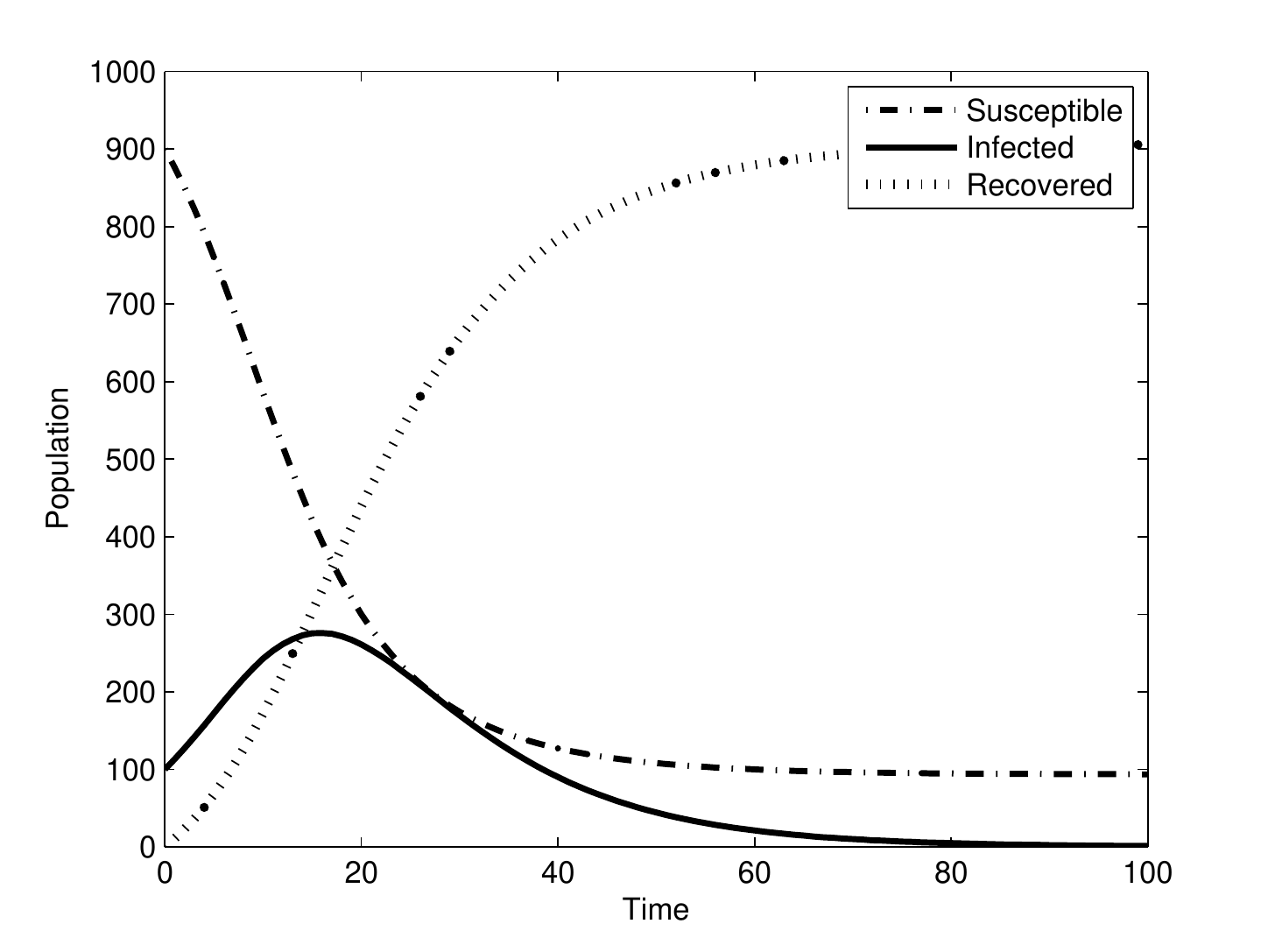}
\caption{ $\beta=0.25$}
\end{subfigure}\\
\begin{subfigure}[b]{0.45\textwidth}
\centering
\includegraphics[scale=0.45]{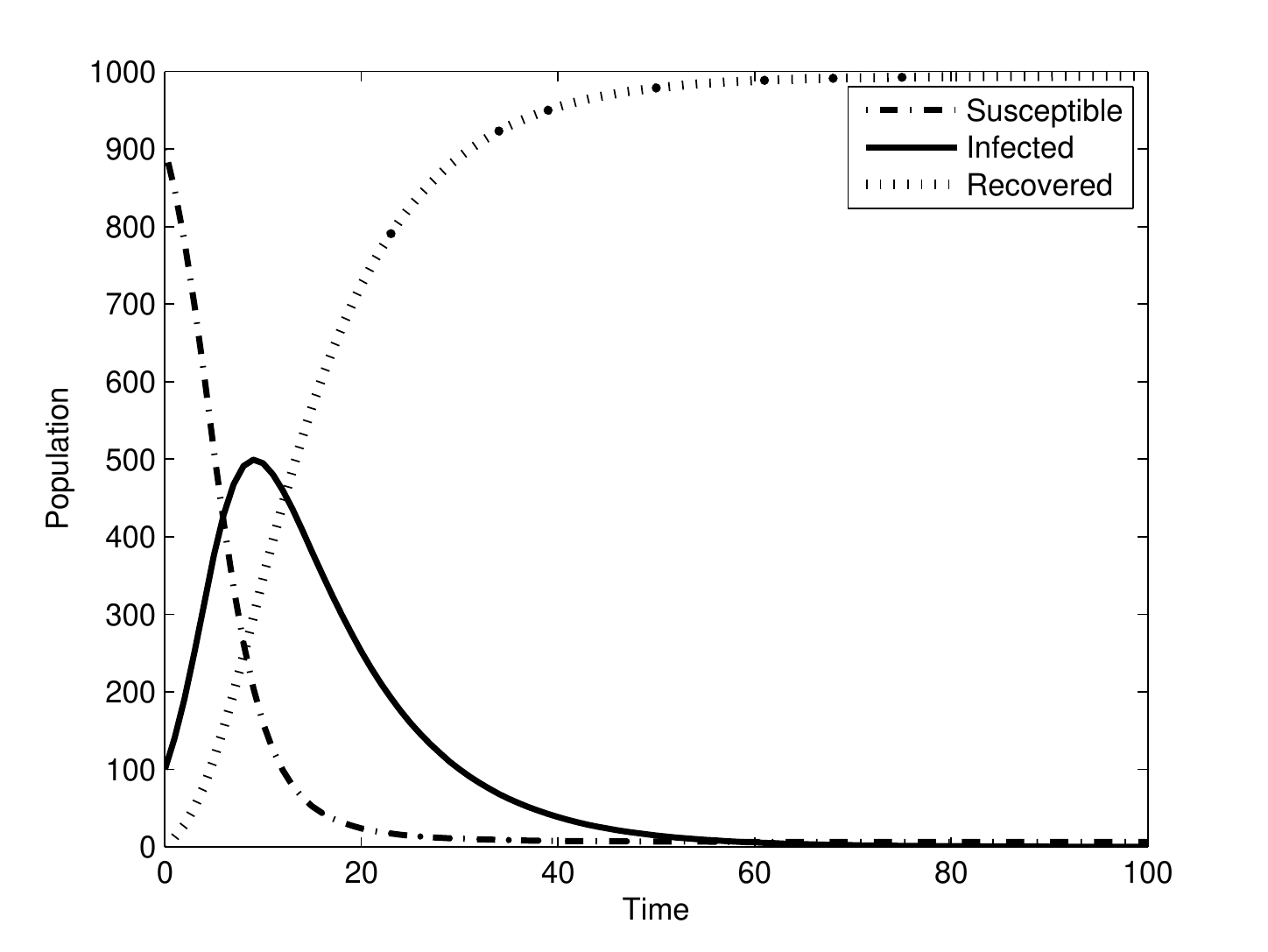}
\caption{$\beta=0.5$}
\end{subfigure}
\begin{subfigure}[b]{0.45\textwidth}
\centering
\includegraphics[scale=0.45]{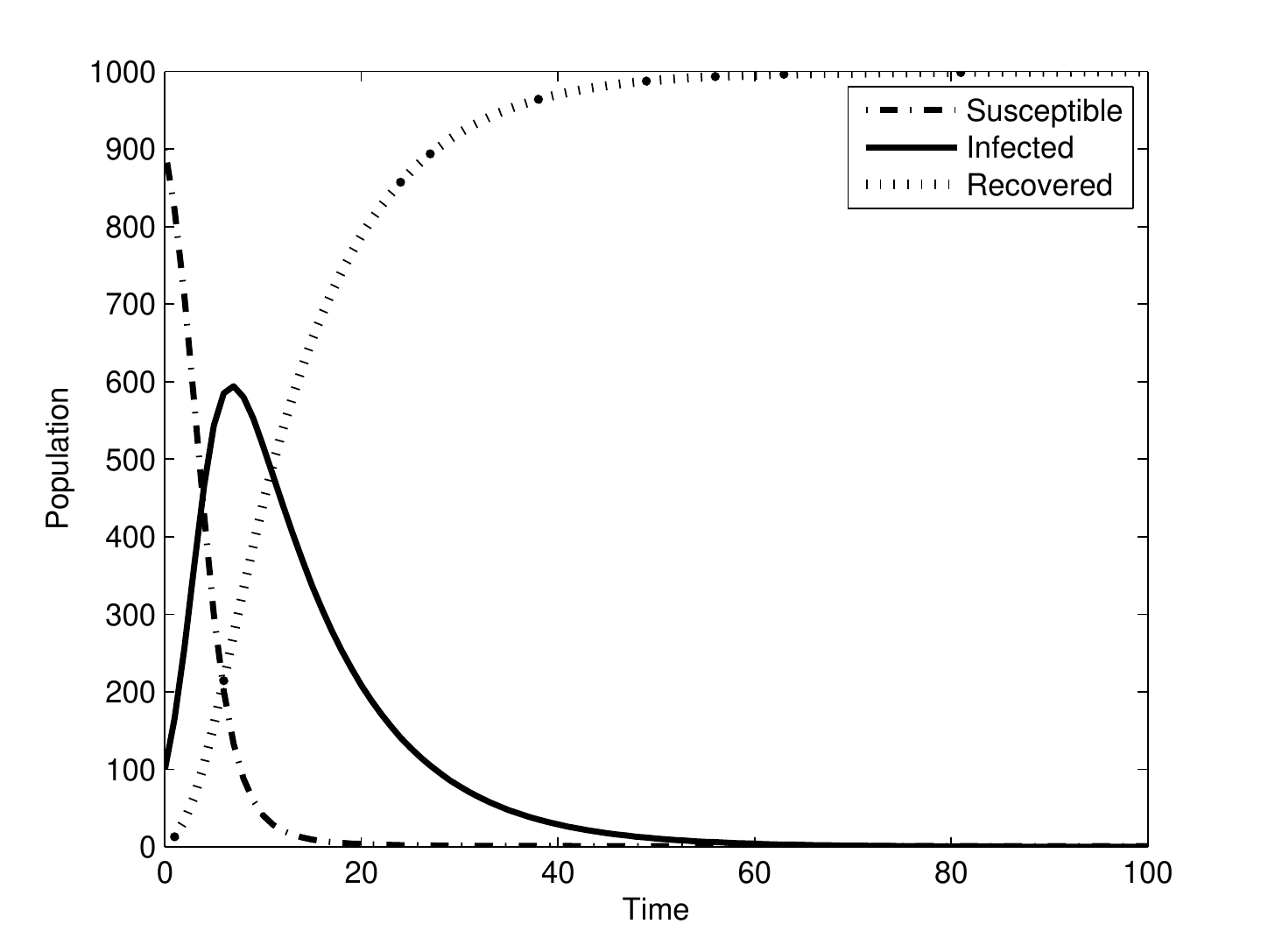}
\caption{$\beta=0.7$}
\end{subfigure}
\caption{SIR model varying infectivity parameter $\beta$ (remaining $\gamma$=0.1)}
\label{beta_analysis}
\end{figure}

Figure \ref{beta_analysis} presents four scenarios for different values of $\beta$,
while the gamma parameter remains constant ($\gamma=0.1$) and the initial values unchanged.
Increasing the value of $\beta$ implies to reach a peak of transmissibility sooner: in
Figure \ref{beta_analysis}b) (for $\beta=0.25$) is close of time 20 reaching about 300 persons,
while in Figure \ref{beta_analysis}d) the highest value is got close to time 10 and infected about
600 individuals. Observing the recovered curve, is possible to visualize that a high value for
infectivity leads to a greater number of target members reached. In the first graphic, the message only
reaches a total of 400 persons of the target audience, approximately. Using a $\beta$ at minimum of 0.5,
the viral campaign achieves almost the target audience in less than 60 units of time.

\begin{figure}
\centering
\begin{subfigure}[b]{0.45\textwidth}
\centering
\includegraphics[scale=0.45]{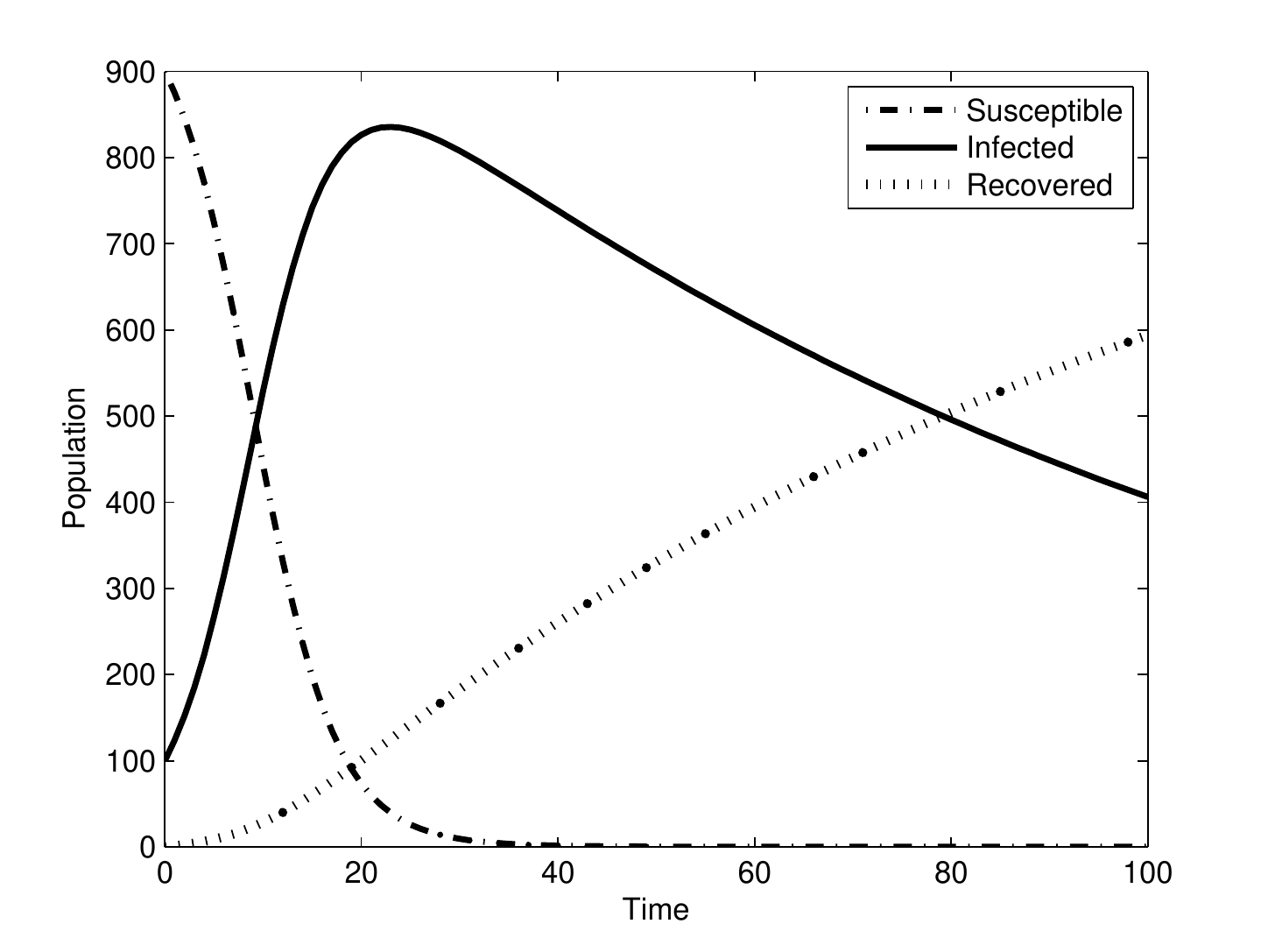}
\caption{$\gamma=0.01$}
\end{subfigure}%
\begin{subfigure}[b]{0.45\textwidth}
\centering
\includegraphics[scale=0.45]{beta025_gamma01}
\caption{ $\gamma=0.1$}
\end{subfigure}\\
\begin{subfigure}[b]{0.45\textwidth}
\centering
\includegraphics[scale=0.45]{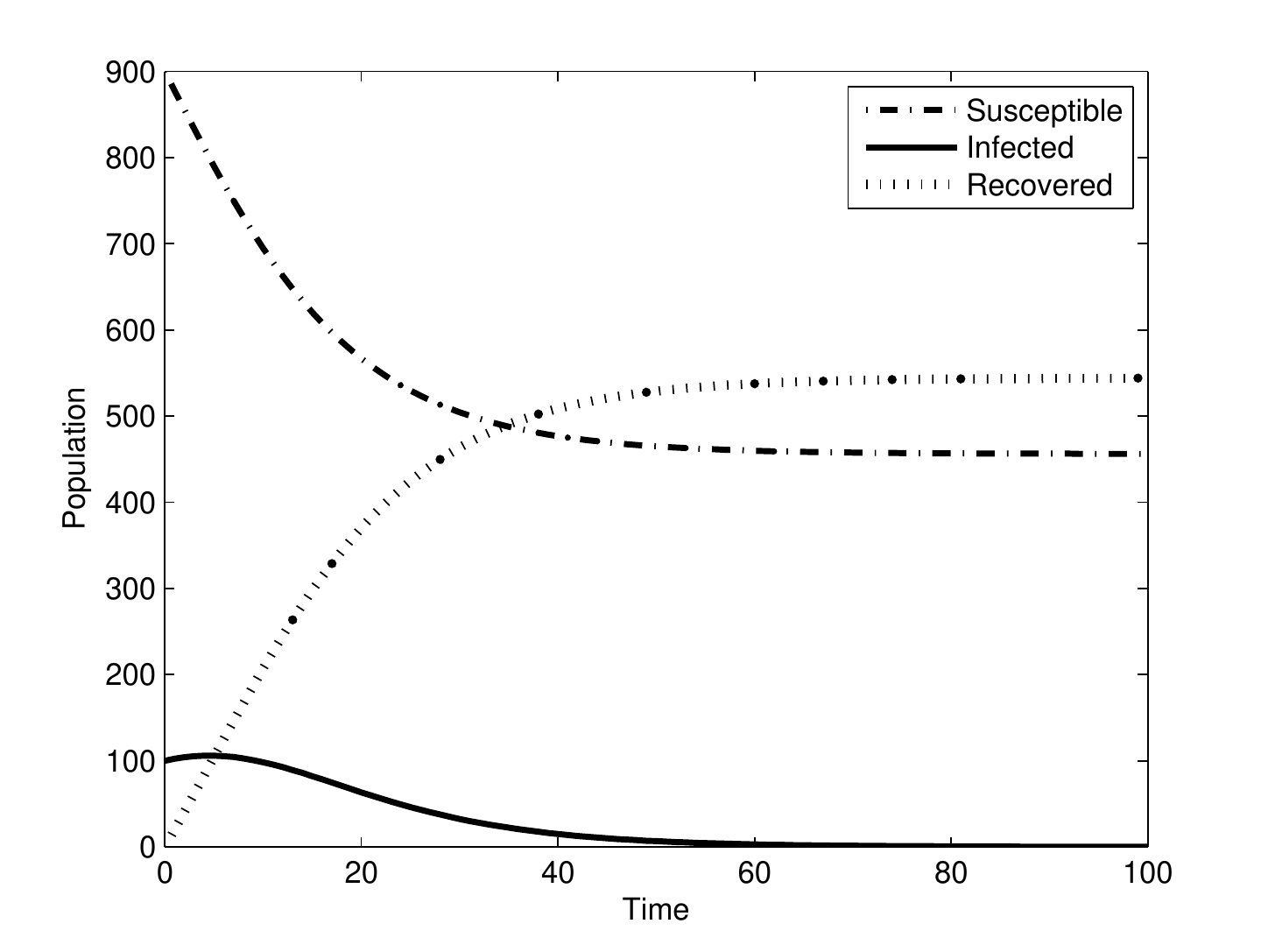}
\caption{$\gamma=0.2$}
\end{subfigure}
\begin{subfigure}[b]{0.45\textwidth}
\centering
\includegraphics[scale=0.45]{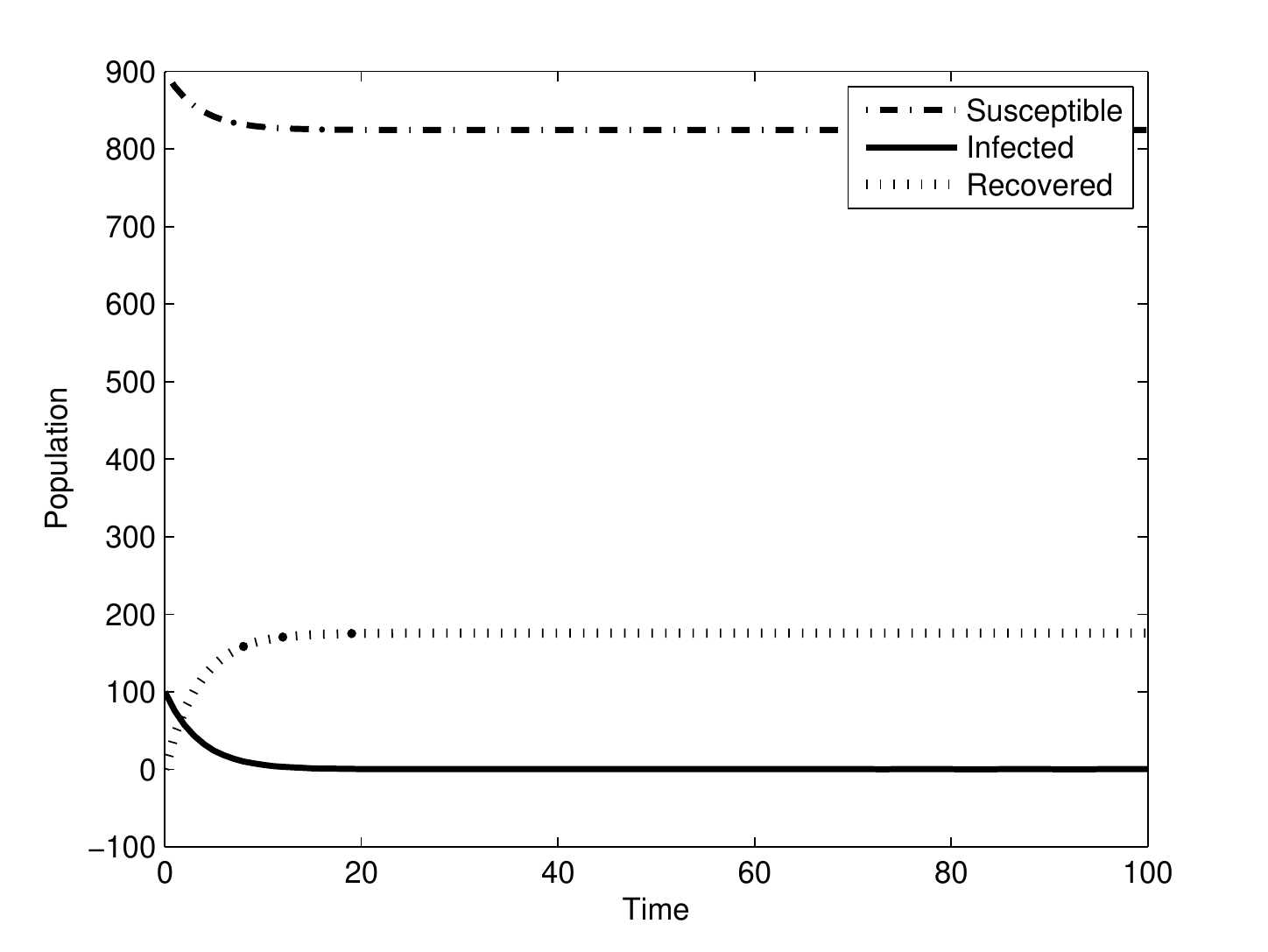}
\caption{$\gamma=0.5$}
\end{subfigure}
\caption{SIR model varying recovery rate parameter $\gamma$ (remaining $\beta$=0.25)}
\label{gamma_analysis}
\end{figure}

Figure \ref{gamma_analysis} is related to the variation of parameter $\gamma$, using the value
0.25 for parameter $\beta$ and remaining the same values for initial conditions. The increasing of $\gamma$
 value leads to a decreasing of sharing message, because an individual tends to
 forget or to be not interested in passing the message, more quickly. In the first graphic ($\gamma=0.01$), all population
is affected for the viral campaign, but in a long period of time; this conclusion can be taken, because in less
than 40 units of time the population ceases to be susceptible. In the last graphic, ($\gamma=0.5$) the total of
population reached by the campaign is less than 200 people.

\subsection{Changing Seed Population }

\begin{figure}
\centering
\begin{subfigure}[b]{0.45\textwidth}
\centering
\includegraphics[scale=0.45]{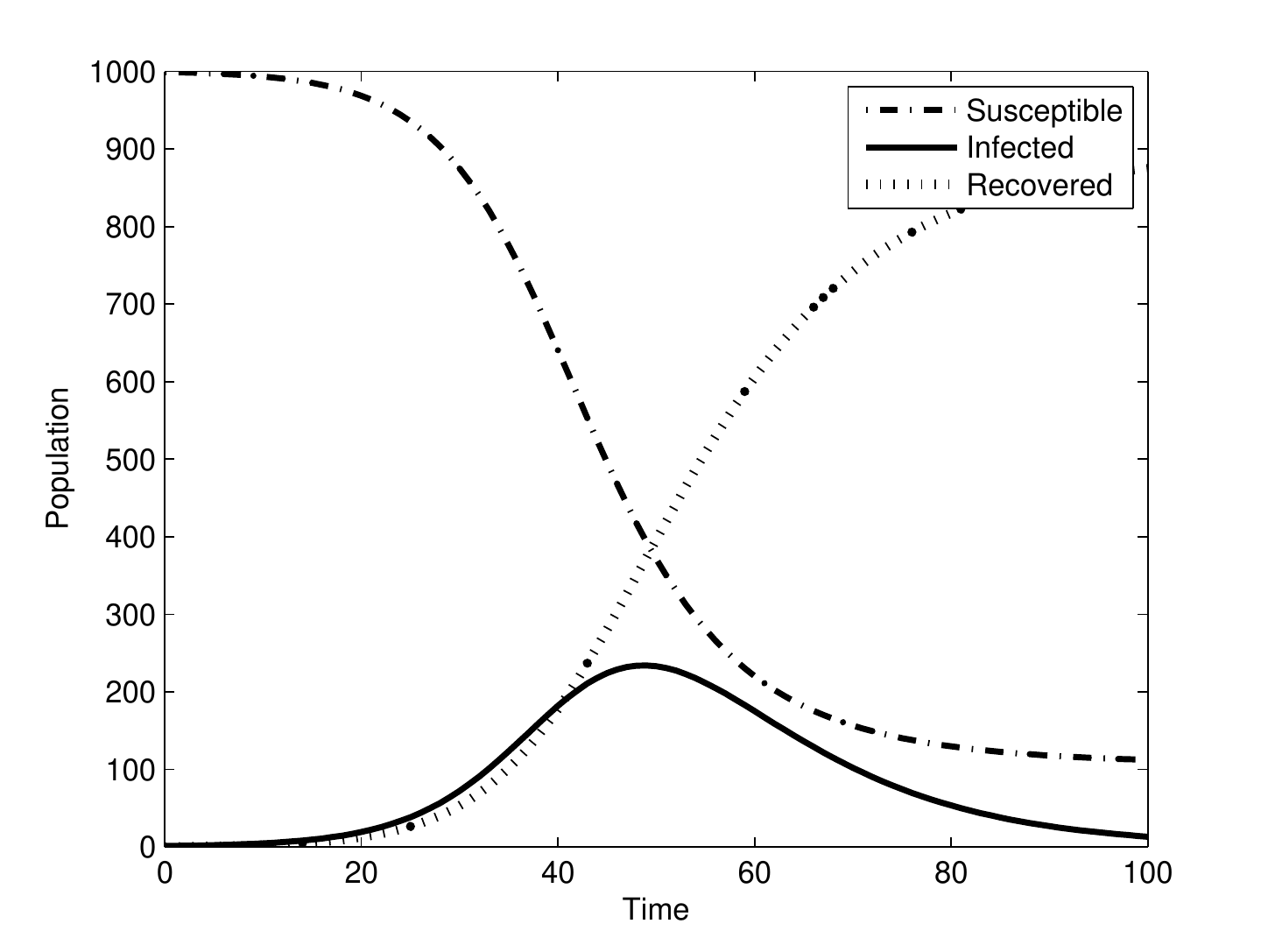}
\caption{$I(0)=1$}
\end{subfigure}%
\begin{subfigure}[b]{0.45\textwidth}
\centering
\includegraphics[scale=0.45]{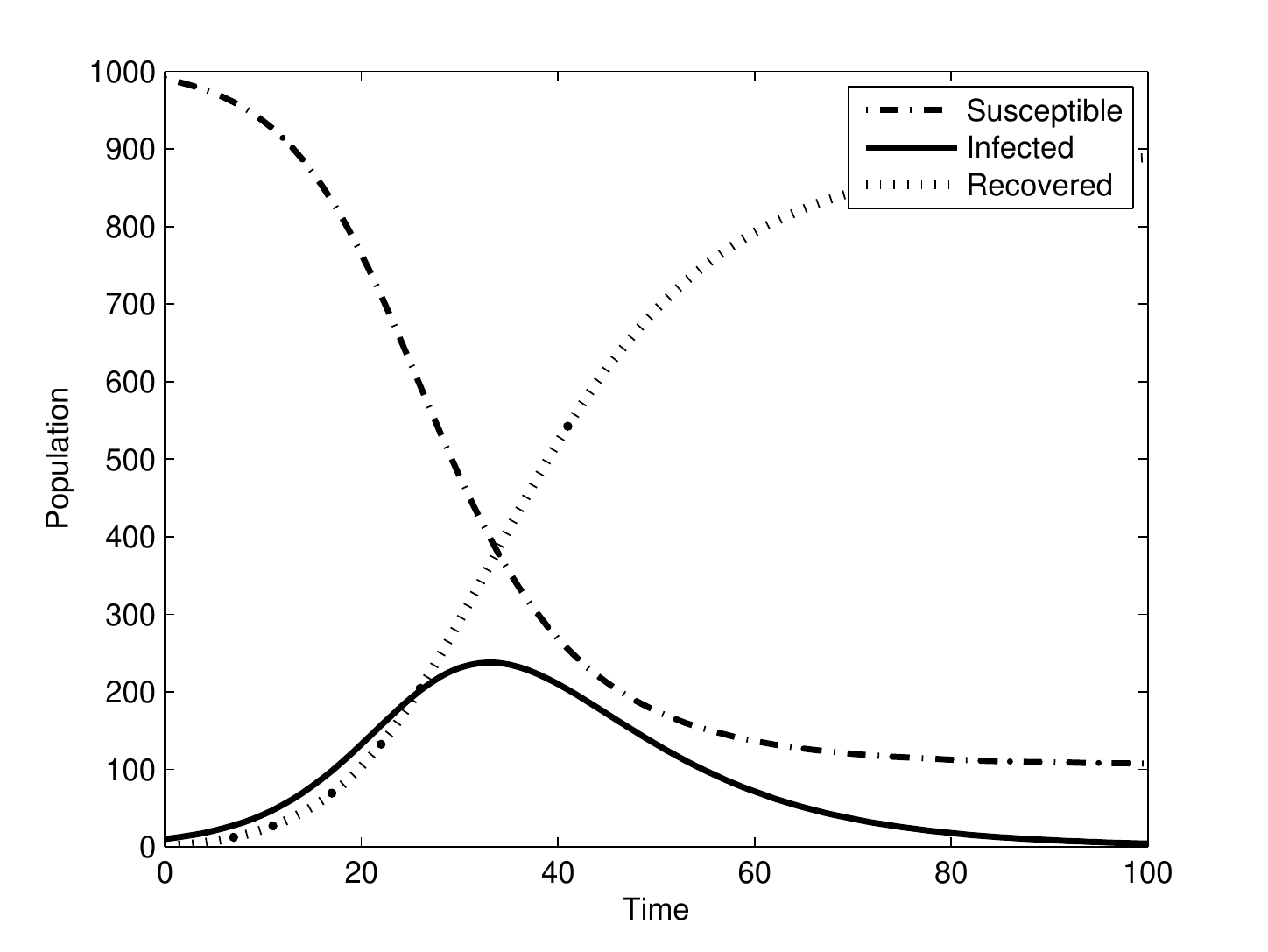}
\caption{ $I(0)=10$}
\end{subfigure}\\
\begin{subfigure}[b]{0.45\textwidth}
\centering
\includegraphics[scale=0.45]{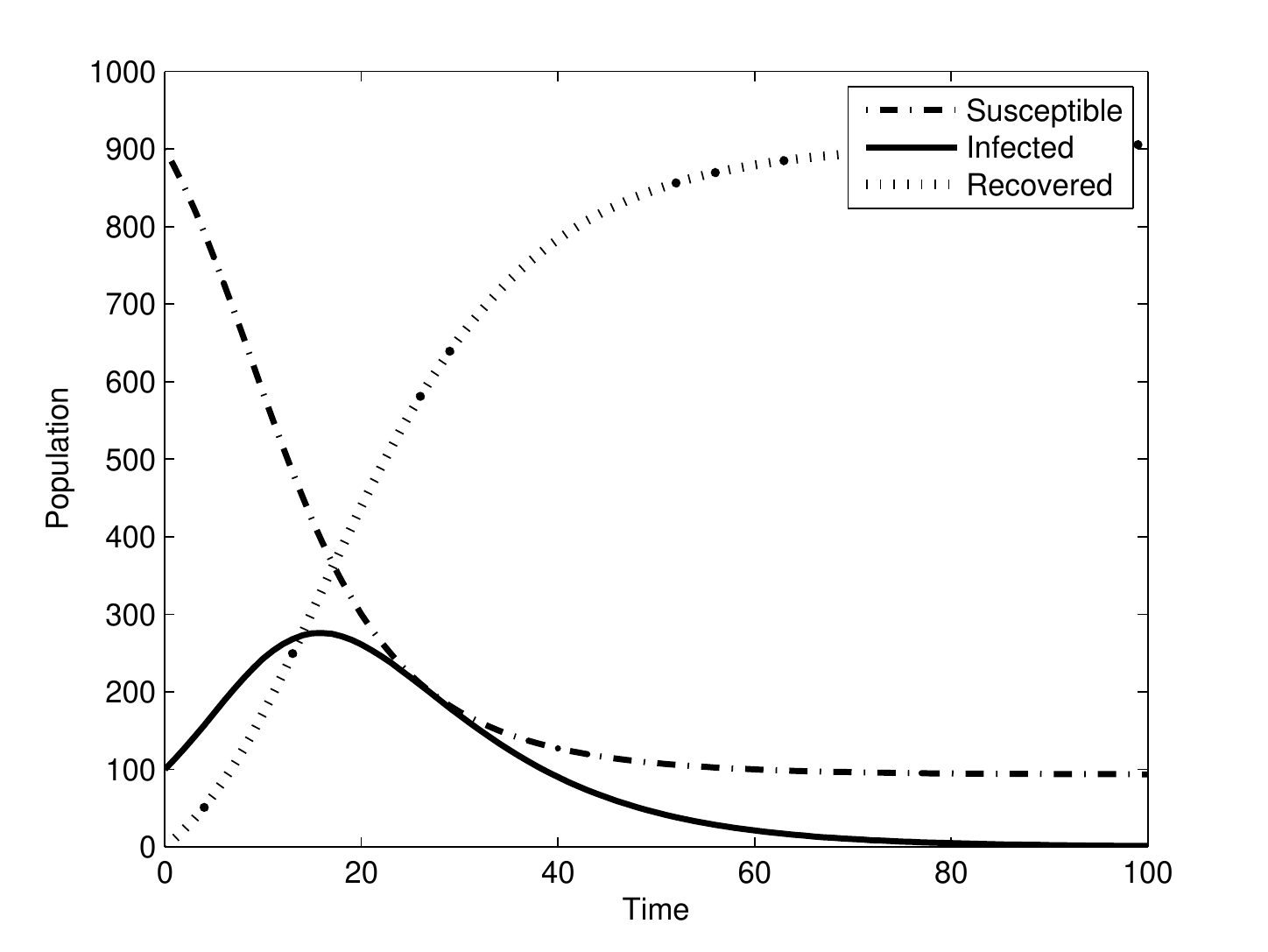}
\caption{$I(0)=100$}
\end{subfigure}
\begin{subfigure}[b]{0.45\textwidth}
\centering
\includegraphics[scale=0.45]{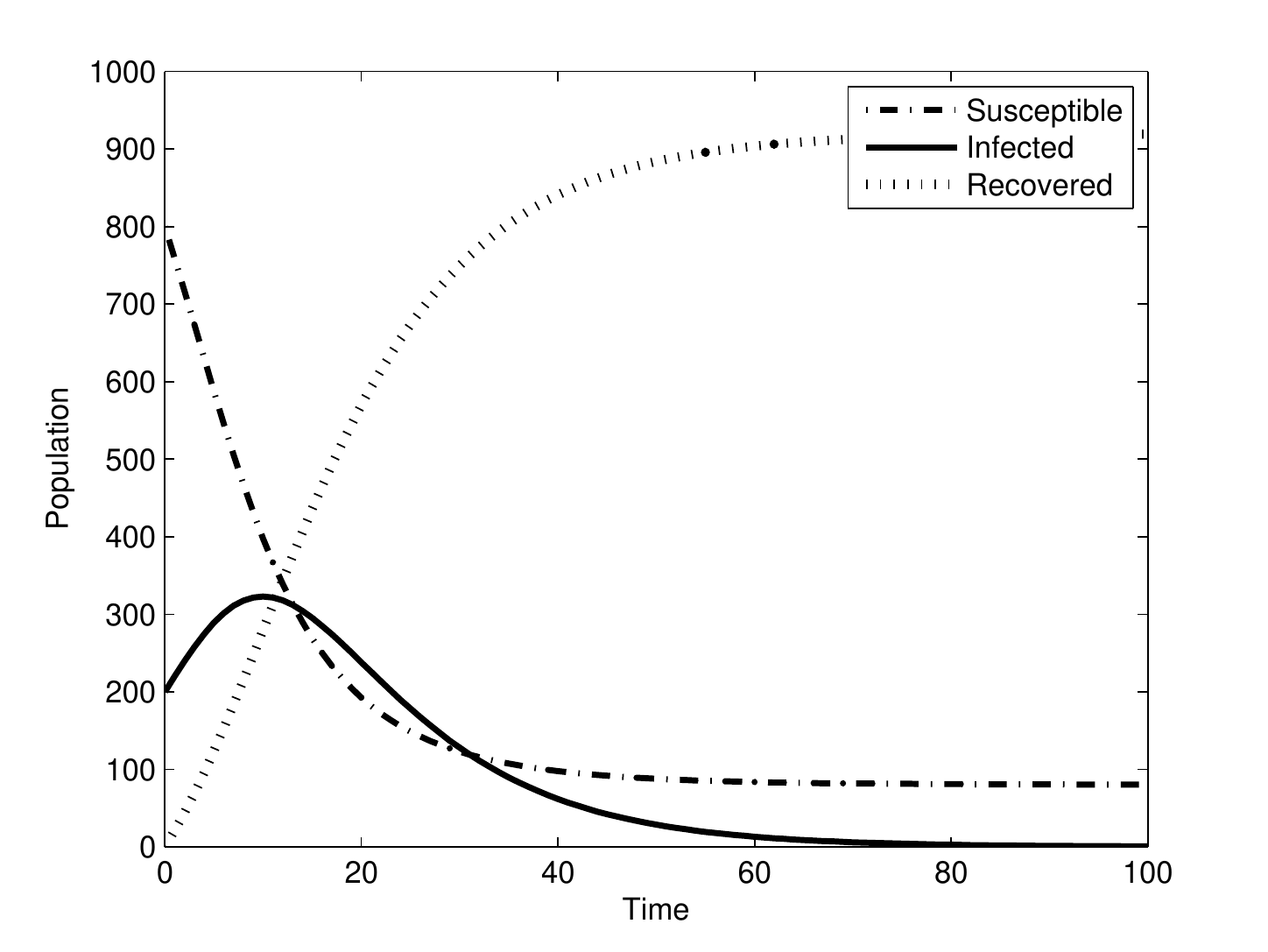}
\caption{$I(0)=200$}
\end{subfigure}
\caption{SIR model varying the initial value of infected individual $I(0)$, remaining $\beta=0.25$, $\gamma=0.1$
and maintaining the population constant, \emph{i.e}, $S(0)=N-I(0)$ and $R(0)=0$}
\label{initial_value_analysis}
\end{figure}

One of the challenges of promoting a VM campaign is to know how many people should
be initially affected, in order to create a huge diffusion of the product or service.
However, increasing the seeding directly may require more costs for the company \cite{Lans2010}.
Figure \ref{initial_value_analysis} reflects the variation of initial value of the target
population that is infected initially. As expected, when the amount of people infected initially is high,
the viral campaign reaches its peak more rapidly. However, the most interesting conclusion is that,
the costs of seeding the campaign to 20\% of the target population produces almost the same effects
(in time and population reached) as invest only in 10\% of the target population to diffuse the campaign.
In most cases, seeding a very large proportion of the population would likely be very costly in
marketing reality, and the results could not be the expected ones.

\section{Conclusions and directions for future work}

This paper presents a model that translate the viral process of a communications marketing campaign.
The parameters used (infectivity and recovery rate) are very sensitivity. The influence of $\beta$ value
is huge in the diffusion of the message: as infectivity increases, the proportion of the
target audience achieved also increases. On other hand, the growth of the $\gamma$ parameter
implies the decreasing of the share of the viral message. In this way, the marketers
should focus their attention to increase the social network (increasing $\beta$)
that has specific characteristics for the commercialization of the product or service,
accurating targeting criteria, in order to
create more interest in each person to passes the message more often, faster and
for a long period of time (decreasing $\gamma$).

The investment in seed the message in an initial set of individuals, must be studied for each case.
For our simulations, to spend more money in the seed population, over 10\%, it is fruitless, because the
viral campaign produce similar effects, but with a higher financial burden for the company.
This measure assumes particular relevance both in an alternative communication strategy,
such as the VM, as in the context of the current economic status, where all investments
in marketing communications should be optimized.

As future work we intend to applied this model to specific viral marketing campaigns, in order to
fit our parameters to the reality.

\section*{Acknowledgements}

This work was partially supported by The Portuguese Foundation for
Science and Technology (FCT), through CIDMA (Center for Research
\& Development in Mathematics and Applications) within
project UID/MAT/04106/2013.

%
%  Bibliography. Follow the usual conventions.
%

\end{document}